# Weighted dynamic finger in binary search trees[*]

John Iacono[†]    Stefan Langerman[‡]

## Abstract

It is shown that the online binary search tree data structure GreedyASS performs asymptotically as well on a sufficiently long sequence of searches as any static binary search tree where each search begins from the previous search (rather than the root). This bound is known to be equivalent to assigning each item $i$ in the search tree a positive weight $w_i$ and bounding the search cost of an item in the search sequence $s_1, \ldots, s_m$ by

$$O\left(1 + \log \frac{\sum_{\min(s_{i-1},s_i) \leq x \leq \max(s_{i-1},s_i)} w_x}{\min(w_{s_i}, w_{s_{i-1}})}\right)$$

amortized. This result is the strongest finger-type bound to be proven for binary search trees. By setting the weights to be equal, one observes that our bound implies the dynamic finger bound. Compared to the previous proof of the dynamic finger bound for Splay trees, our result is significantly shorter, stronger, simpler, and has reasonable constants.

## 1 Introduction

In spite of defining one of the most fundamental classes of data structures in Computer Science, the full power of binary search trees (BSTs) is still not fully understood. The discovery of self-adjusting trees such as *Splay trees* [17], which can perform sequences of searches faster by adapting to their distribution, prompted what is probably the most tantalizing question in the field: How well can BSTs adapt, and in particular can a generic algorithm perform as well on a sequence of searches as an algorithm that would be specifically tailored to that sequence?

**BST model.** Formally, let $S = s_1, s_2, \ldots s_m$ be a sequence of $m$ searches[1], where each of $s_i \in [1..n]$. The BST model has been formalized by requiring a search to be implemented with single pointer that starts at the root, and which at unit cost can be moved to one of the children, the parent, or perform a single rotation, with the obvious requirement that to successfully execute the search the pointer must touch the node containing the searched item at some point during the execution of the search. This model encompasses what one normally thinks of as a BST-model algorithm. Let $\text{OPT}(S)$ be the fastest any binary search tree can execute the search sequence $S$ in the BST model given a choice of the initial tree[2]. Although $\text{OPT}(S)$ is well defined, it is unknown whether it is efficiently computable.

---

[*]This work appeared as [13].

[†]Algorithms Research Group, Département d'Informatique. Université Libre de Bruxelles, Bruxelles, Belgium and Department of Computer Science and Engineering, Tandon School of Engineering, New York University, 2 MetroTech Center 10FL, Brooklyn NY 11201. Research supported by NSF grants CCF-1319648, CCF-1533564 and MRI-1229185, a Fulbright Research Fellowship and from the Fonds de la Recherche Scientifique-FNRS under Grant no MISU F 6001 1.

[‡]Directeur de recherches du F.R.S.-FNRS. Algorithms Research Group, Département d'Informatique, Université Libre de Bruxelles, Bruxelles, Belgium.

[1]Here only searches are considered, not insertions or deletions, so w.l.o.g. it is assumed the BST stores the integers $[1..n]$.

[2]The choice of initial tree does not asymptotically impact $\text{OPT}(S)$ if $m = \Omega(n)$ since any tree can be converted into any other tree in $O(n)$ time in the BST model [18].

**Dynamic optimality conjecture.** What has made the field of binary search trees interesting is the *dynamic optimality conjecture* [17] which posits the existence of online binary search tree algorithms that execute all sufficiently long $S$ in time $O(\text{OPT}(S))$; a BST algorithm that satifies this conjecture is said to be *dynamically optimal*. It must be emphasized that there are no distributional assumptions in this conjecture. To satisfy it a BST algorithm must execute all sequences as fast as any algorithm that has foreknowledge of the sequence of searches and infinite processing time to determine which are the best unit-cost BST operations to execute that sequence. So far there are two serious contenders to satisfy this conjecture. Most famously is Splay trees [17]. However, there is an another BST algorithm known as *GreedyASS* [14], which was originally stated as a simple greedy offline algorithm but has been shown to have an online equivalent [6]. Neither GreedyASS nor Splay trees are known to be dynamically optimal. A BST algorithm was proposed in [12] that is dynamically optimal if any BST algorithm is, however the algorithm is spectacularly impractical and requires a superexponential amount of non-BST-model work to determine what unit-cost BST operations to execute next. A BST-model structure called *Tango trees* was introduced in [7] and runs in time $O(\text{OPT}(S) \log \log n)$; unfortunately for some classes of sequences $S$, this is tight and the general method seems to not allow further improvement. Tango trees also lack the simplicity and elegance of Splay trees and GreedyASS.

**Bounds.** Several non-trivial upper bounds for Splay trees and GreedyASS are known. The working-set bound, which is based on temporal locality (items searched recently are fast to search) is known for Splay trees [17] and GreedyASS [10]. It states that the amortized time to execute $s_i$ is proportional to the logarithm of the number of distinct items searched since $s_i$ was last searched. The working set bound has been shown [11] to imply several other bounds such as static finger bound, the static optimality bound, as well as $O(\log n)$ amortized time which were originally presented separately. This was shown to be the best possible bound when the proximity of key values is not taken into effect [6].

The dynamic finger bound is based on spatial locality (a search is fast if its key value is close to the previous search). It states that the amortized time to execute $s_i$ is $O(\log |s_i - s_{i-1}|)$. This was proven for Splay trees in a two-volume work [5, 4], but remained open (until this work) for GreedyASS.

Neither the working set bound nor the dynamic finger bound imply each other, and both are easily shown to not be tight on some classes of search sequences. There has been a bound introduced, the unified bound, that does imply both dynamic finger and working set; informally it requires a search to be fast if it is close in time to something close in keyspace. However, no BST-model data structure is known to have the unified bound (one highly-engineered structure was claimed [9] but later called into question [16]).

There are also non-trivial lower bounds on the time it takes to execute deterministic search sequences in the BST model. Wilber [20] produced two bounds, neither of which are known to imply each other. Those have been improved to the bounds of [6] which imply the Wilber bounds and are also not known to not be tight.

**Lazy finger.** This work presents an upper bound which is an elegant generalization of the dynamic finger bound and the static optimality bound and proves that this bound holds for GreedyASS. A BST algorithm $\mathcal{A}$ has the static finger bound if for any fixed tree $T$ the time to execute a sufficiently long search sequences on $\mathcal{A}$ is asymptotically the same as on $T$; this bound is related to the entropy of the frequencies of searching each item and is known to hold for Splay trees [17] and GreedyASS [10]. A generalization of this bound, which we introduced in [2] and called the *lazy finger bound*, also assumes a fixed tree $T$ but instead of measuring the cost from the root, measures it from the previous search. This is a much stronger bound, and is not related to entropy.

One generalization of the dynamic finger bound comes from weighted random search trees [15]. The idea is to give each item $i$ a weight, and to have a search be fast if the weights of the previous and current searches are large compared to the sum of the weights between the two items. Formally, a search structure has the *weighted dynamic finger bound* for any set of positive weights $w_i$ if the cost to execute search $s_i$ is

$$O\left(1 + \log \frac{\sum_{\min(s_{i-1}, s_i) \leq x \leq \max(s_{i-1}, s_i)} w_x}{\min(w_{s_i}, w_{s_{i-1}})}\right)$$

amortized. Although random weighted BST have this bound for a specific set of weights, before this work, no search structure, in the BST model or elsewhere, was known to have the weighted dynamic finger bound without being provided the weights. Note that setting the weights to be equal gives the dynamic finger bound. This bound is easily seen to be stronger than the dynamic finger bound. For example, consider the sequence of searches $S = \sqrt{n}, 2\sqrt{n}, \ldots \sqrt{n}\sqrt{n}, \sqrt{n}, 2\sqrt{n}, \ldots, \sqrt{n}\sqrt{n}, \ldots$, where we assume $n$ is a prefect square. The dynamic finger bound bounds each search as $O(\log n)$ as each search is to a key value $\sqrt{n}$ from the previous one giving a bound of $O(\log |s_i - s_{i-1}|) = O(\log \sqrt{n}) = O(\log n)$. The working set bound also bounds this sequence as taking $O(\log n)$ as there are generally $\sqrt{n}$ distinct items searched between searches of an item. However, by setting the weights of the $\sqrt{n}$ searched items to 1 and the weights of the other items to $1/n$ the weighted dynamic finger would bound the search cost to be $O(1)$ amortized.

The central result of [2] is that for any sequence $S$, if you pick the $T$ that minimizes the lazy finger bound you get a bound asymptotically equivalent to the weighted dynamic finger bound. The main result of this paper is to show that these two equivalent bounds hold for GreedyASS. In doing so we obtain a bound for that data structure that is better than the dynamic finger bound, making it a plausible candidate for dynamic optimality. This is significant as the dynamic finger bound has been the main impediment in producing better bounds for search tree algorithms and has stood unimproved for any plausible contender for dynamic optimality until now, a span of 25 years since it first appeared. The proof of the dynamic finger bound for Splay trees spans two volumes [5, 4] and is a tour-de-force of various techniques, observations, and inverse Ackermanns. It is very complicated, lengthy (85 pages), and has enormous constants (the leading constant is 42,000 and the constant on a lower-order log-log-times-inverse-Ackermann term is $10^{16}$). Proving a bound that implies the dynamic finger bound, would require one to try to extend the proof there, or to take an entirely new approach. We have opted for the latter. Our proof is a completely novel approach that yields a relatively straightforward amortization with respect to how the algorithm GreedyASS runs and the reference tree $T$ which we are trying to prove lazy-finger competitiveness with. Given the lazy finger bound our results from [2] immediately give the weighted dynamic finger bound. The resulting argument is fairly simple, has low constants, and perhaps most tantalizingly leaves room for possible improvements.

## 2 Binary search trees, the geometric view, and GreedyASS

In [6] a simple geometric view of BST algorithms was introduced. Given a BST algorithm executing a sequence of $m$ searches $s_1, \ldots, s_m$ on a BST storing the integers $[1..n]$, the geometric view consists of a set of points on the $n \times m$ grid, where a point $(i, j)$ is in the geometric view iff the node containing $i$ was touched in the $j$th search $s_j$. Consider the following property: Given a set $P$ of points on a grid, two points not on the same row or column are said to be *arborally satisfied*, if the rectangle defined by these two points contains a third point. Two points on the same row or column are always considered arborally satisfied as well. A set of points where every pair of points is arborally satisfied is called an *Arborally Satisfied Set (ASS)*. It was shown that the geometric view of a BST algorithm when run on any search sequence is an ASS, and furthermore, for any ASS there is a BST execution that produces that set. Thus ASS sets completely characterize BSTs, and do so in a way that is simple and in particular exempts one from the difficult process of reasoning directly about rotations.

Clearly, in order to execute the $j$th search the algorithm must touch $s_j$ at time $j$; in the geometric view this corresponds to including the point $(s_j, j)$ in the set. Thus any BST algorithm executes a sequence of searches $s_j$ in the geometric view if and only if (1) it contains the *search points* $P(S) = \{(s_j, j) | j \in [1..n]\}$ and (2) it is an ASS. Let $\text{OPT}(S)$ be the size of a minimal-size ASS superset of $P(S)$. If one can find an ASS superset of the search points $P(S)$ of size $O(\text{OPT}(S))$, then one has found an asymptotically optimal way of executing $S$ in the BST model. Tango trees [7] find a set of size $O(\text{OPT}(S) \log \log n)$; no polynomial-time method of finding an ASS superset of size $o(\text{OPT}(S) \log \log n)$ is known.

It is easy to formulate a greedy algorithm to compute an ASS superset of $P(S)$ in the geometric view, which has been called GreedyASS. This method starts with $P(S)$ and performs a vertical line sweep, adding points to make the set an ASS. At each step the algorithm maintains the invariant that the point set below the sweep line is an ASS. When the sweep line is at row $i$, any ASS violations are fixed by adding the minimal number of points to row $i$ to obtain the ASS property. These violations occur when $(i, s_i)$ and a point below row $i$ define a rectangle that violates the ASS property; the minimum way to remove these ASS violations is by placing points on the upper corners of the rectangles that is not $(i, s_i)$. See Figure 1 for a worked-out example of a GreedyASS execution in the geometric view.

This algorithm was shown in [6] to be equivalent to the following algorithm in the tree view first proposed by Lucas [14]: search for $s_k$ and then reconfigure the search path (only) to have the next search as high as possible and recurse on any remaining elements of the search path. This algorithm, as stated, is not online, and in fact can be seen as the natural offline greedy algorithm. However, it was shown in [6] that there is an online algorithm with equivalent asymptotic cost.

## 3 Main result

### 3.1 Overview of method and notation

Let $S := s_1, \ldots s_m$ be a sequence of searches executed by GreedyASS and let $P_i$ be the sets of points in the geometric view of GreedyASS executing the first $i$ operations of $S$; $P := P_m$. The *reference tree* $T$ is an arbitrary *leaf-oriented* binary search tree, that is, a tree whose leaves are labeled $1, 2, \ldots, n$ left-to-right and whose internal nodes are unlabeled. Our goal is to show that GreedyASS will execute $S$ as fast as the static tree $T$ would with a lazy finger, asymptotically. Let[3] $d(x, y) := d_T(x, y)$ be the distance in the tree $T$ from $x$ to $y$ via their LCA, measured in nodes. Thus we wish to show that GreedyASS has a cost of $O(d(s_{i-1}, s_i))$ amortized to execute the $i$th search, and we do so by establishing a potential function that is parameterized by $T$.

We emphasize that the running time of GreedyASS is independent of $T$. Our analysis is based on viewing GreedyASS exclusively in the geometric view. Thus any time we mention tree notation it refers to the reference tree $T$ and *not* GreedyASS viewed as a BST algorithm. A *point* will always refer to an element the geometric view, and a *node* will always refer to a node of $T$. To avoid confusion, $i, j, k$ will be used to refer to the vertical time axis in the geometric view, and $x, y, z$ will refer to the horizontal key axis. In the geometric view, *left* and *right* are unambiguous, and we adopt the convention in our figures that *up* corresponds to increasing time (the *future*) and *down* decreasing time (the *past*).

We refer to the integers $[1..n]$ as *items*. We distinguish between the terms *accessed* and *searched*. Item $x$ is *searched* at time $i$ if $s_i = x$. Item $x$ is *accessed* at time $i$ if $(x, i) \in P_i$. Node $x \in [1..n]$ refers to the leaf node containing $x$ in the tree. The most recent access of $x$ at time $i$, $\text{MRA}_i(x)$ is the highest point in $P_i$ in column $x$. To get the individual coordinates of a point, we use $.x$ and $.time$; e.g. $\text{MRA}_i(x) = (\text{MRA}_i(x).x, \text{MRA}_i(x).time)$.

For a node $v$ of $T$, let $T_v$ be the subtree of $T$ rooted at $v$, and let the *interval of $v$*, $I_v$ be the range spanned by the labels of the leaves in $T_v$. The *reference forest* at time $i$, $F_i$ is obtained by removing all ancestors of $s_i$ from $T$, see Figure 2. The root of item $x$ at time $i$, $r_i(x)$ is the root in $F_i$ that contains $x$ in its subtree.

### 3.2 Potential

For every time $i$, a nonzero *item potential* $\varphi_i(x)$ is assigned to each of the items $x$ that have been accessed so far, and each root $r$ of the reference forest $F_i$ will be assigned a *virgin potential* denoted as $v_i(r)$. The potential of the structure at time $i$, $\Phi_i$ will simply be the sum of the item potentials of each item in $[1..n]$ and the virgin potentials of each root in $F_i$. We now formally define how these constituent potentials are computed.

At time $i$ each item $x$ is assigned one of three *location classes*. We present the formal definitions here but note that they have a simple geometric interpretation which is illustrated in Figure 3. Let $\text{ROW}_i := \{x | (x, i) \in P_i\}$ be the set of all items accessed at time $i$. Let $\text{L-INT}_i(x)$ and $\text{R-INT}_i(x)$ be the points in $P_i$ to the left and right of and on the same row as the most recent access of $x$, $\text{MRA}_i(x)$, if they exist. That is, $\text{L-INT}_i(x) := \max\{y | y < x, y \in \text{ROW}_{\text{MRA}_i(x).time}\}$ and $\text{R-INT}_i(x) := \min\{y | y > x, y \in \text{ROW}_{\text{MRA}_i(x).time}\}$

**External.** An item $x$ is *i-left external* if there are no accesses to the left of $\text{MRA}_i(x)$ and thus $\text{L-INT}_i(x)$ is undefined. Observe that for $i$-left external $x$ there are no points in $P_i$ other than $\text{MRA}_i(x)$ in the quadrant at or above and to at or to the left of $\text{MRA}_i(x)$; such an access would violate the ASS property or the definition of left external. The definition of $i$-right external is symmetric and a node is $i$-external if it is $i$-left or $i$-right external.

**Glancing.** An item $x$ is *i-left glancing* if it is not $i$-external and no points $P_i$ are in the quadrant strictly above and at or to the left of $\text{MRA}_i(x)$. The definition $i$-right glancing is symmetric and an item is $i$-glancing if it is $i$-left or $i$-right glancing.

**Internal.** An item that is neither $i$-external nor $i$-glancing is $i$-internal.

For the purpose of the analysis, at each time $i$ we will assign a *color* to each item $x$ and the point $\text{MRA}_i(x)$. We refer to the points in $P_i$ that are the $\text{MRA}_i(x)$ of some $x$ as *active* at time $i$, they are the ones which are colored. Each of these colors denotes a different potential function. First we describe the four potential functions, and then describe how the color of each item, and therefore its potential, is determined. For $i$-external elements $x$, let L-Ext$_i(x)$ and R-Ext$_i(x)$ be the $i$-external items closest to the left and right of $x$, if they exist (among the sorted order of all $i$-external elements.) These values are well defined for all $i$-external $x$ except the first and last one, respectively.

---

[3] We assume $S$ and $T$ are fixed and allow further notation (such as $P$) to be defined without explicitly indicating a dependence on $S$ and $T$.

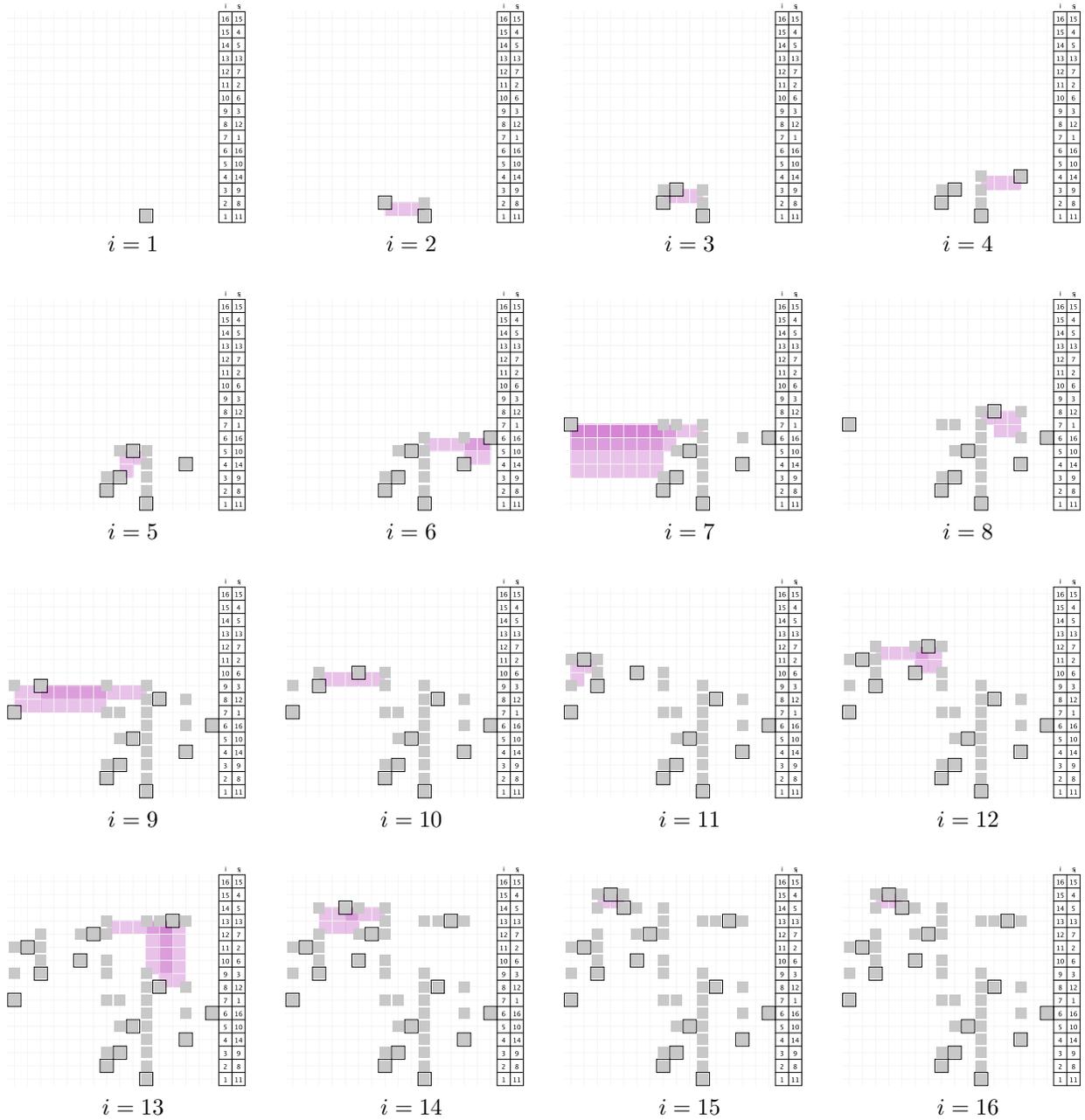

Figure 1: Illustration of the GreedyASS algorithm. The points with the black squares represent the searches $s_i$. The purple shaded rectangles represent the unsatisfied rectangles which GreedyASS satisfies by placing a point at the upper non-$s_i$ corner.

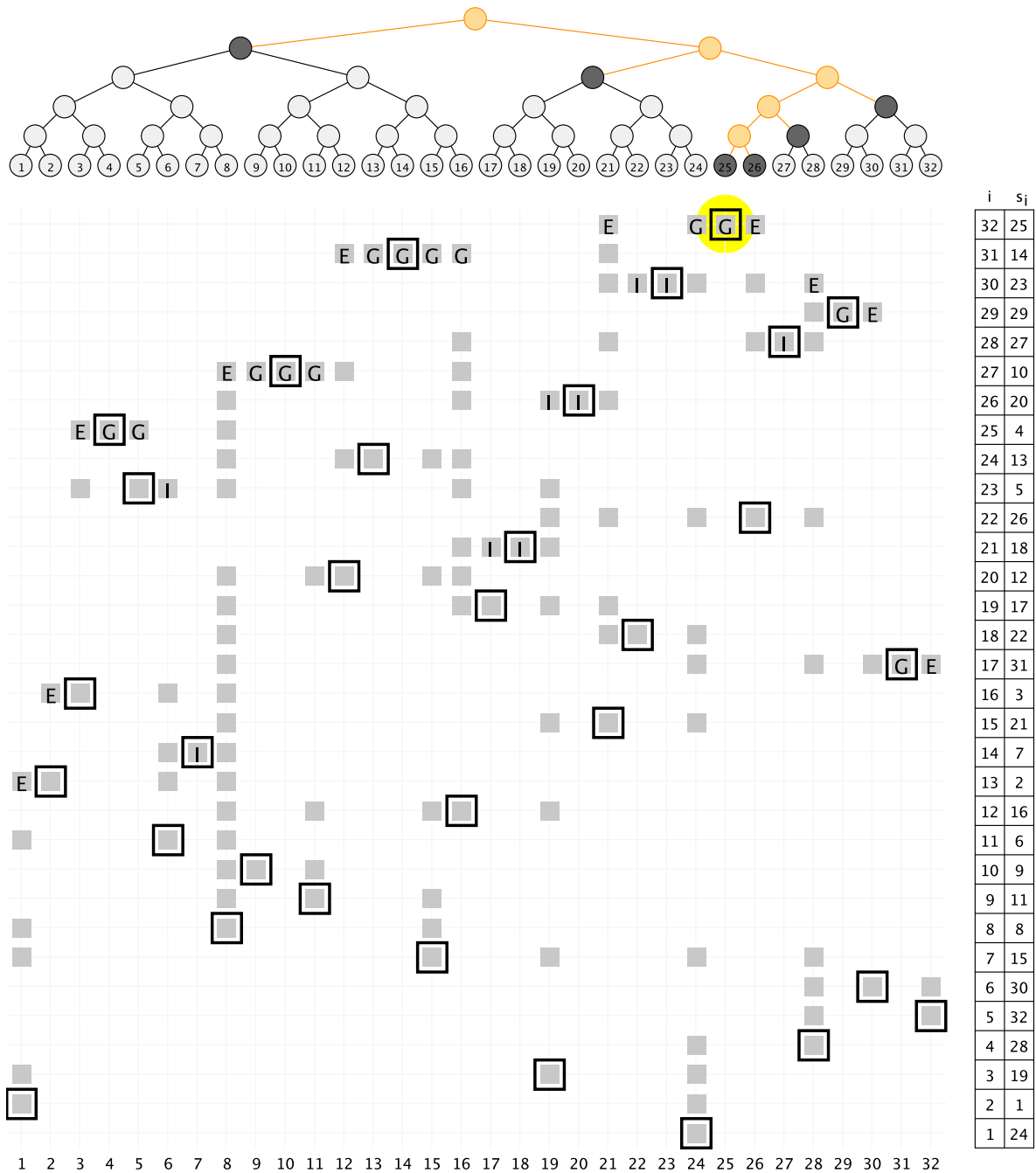

Figure 2: The reference tree $T$ and forest $F_i$. Here we illustrate a reference tree $T$ that is leaf-oriented over the integers 1..32. We could have chosen any such tree, but for simplicity we picked a balanced tree. The reference forest is obtained by noting that the most recent (top) search is to 25, and removing all of the ancestors of the leaf labeled 25 from $T$. We use orange to illustrate those nodes that have been removed from $T$ to make $F$. The roots of $F$ are colored dark gray.

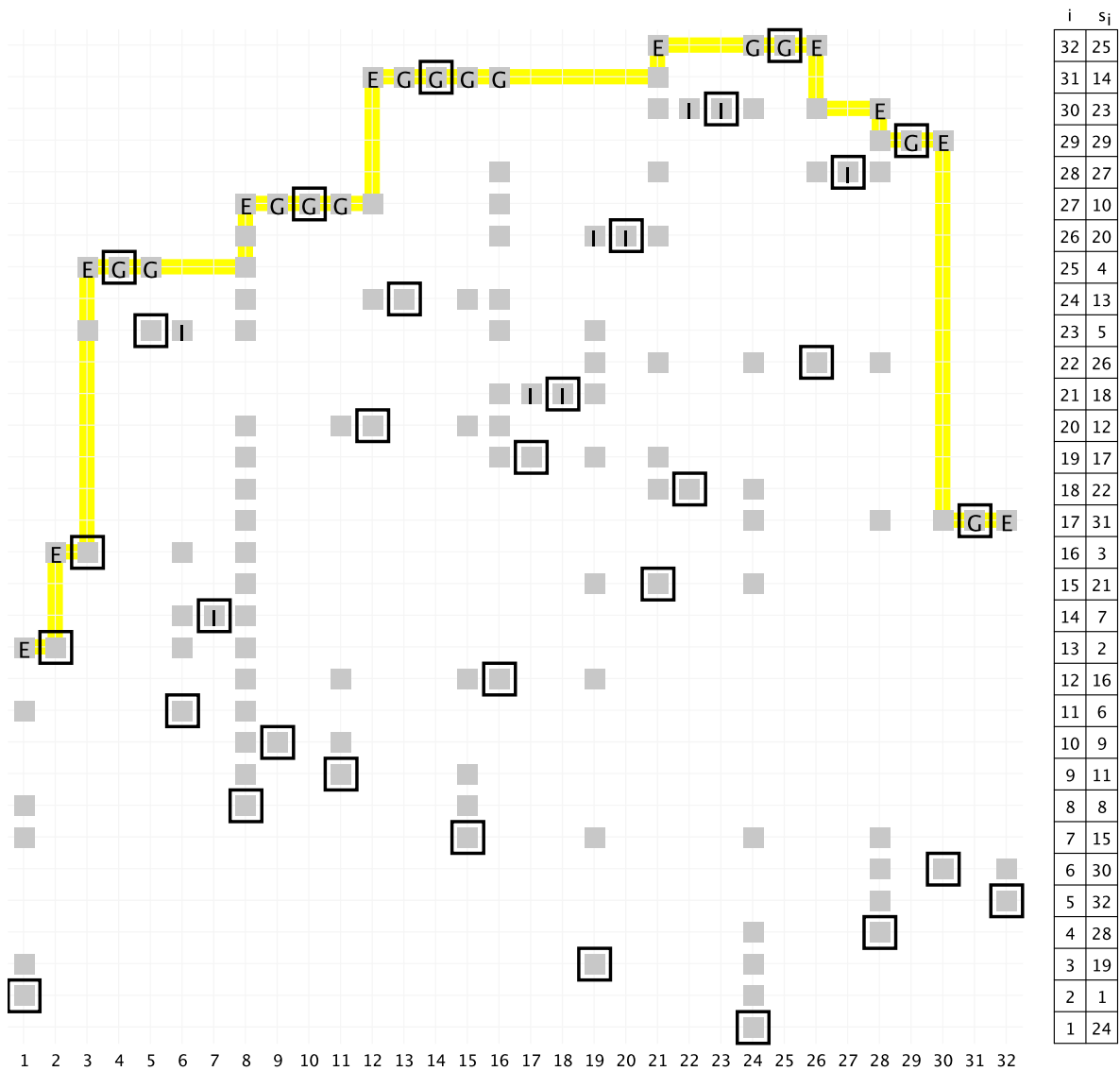

Figure 3: Item types. The highest point in each column is classified as internal, external, or glancing, indicated on the diagram with a single letter. Graphically, these categories are based on the yellow upper orthogonal convex hull of the points. The corners on the yellow are external, the non-corners on the yellow are glancing and the rest are internal.

**Red.** The red potential of $x$ at time $i$ is Red$_i(x) := d(x, \text{Lca}(\text{L-Int}_i, \text{R-Int}_i))$. This potential is only well-defined when L-Int$_i(x)$ and R-Int$_i(x)$ exist, and thus is only well-defined when $x$ is $i$-internal or $i$-glancing. See Figure 6.

**Green.** The green potential of an $i$-external element $x$ at time $i$ is Green$_i(x) := d(x, \text{Lca}(\text{L-Ext}_i(x), \text{R-Ext}_i(x))) + 1$. When $x$ is the first or the last $i$-external element, we set Green$_i(x) := \infty$. See Figure 5.

**Blue.** Let $r_i(x)$ be the root in the reference forest $F_i$ that contains $x$. The blue potential of $x$ at time $i$ is Blue$_i(x) := d(x, r_i(x)) + 1$. See Figure 4.

**Purple.** The purple potential of $x$ at time $i$ is the highest the blue potential has been since its most recent access, Purple$_i(x) := \max_{j=\text{Mra}_i(x).time}^{i} \text{Blue}_j(x)$.

The color of an item $x$ at time $i$, denoted as Color$_i(x)$, is a function Color $: [1..n] \mapsto \{\text{Red}, \text{Green}, \text{Blue}, \text{Purple}\}$. The color determines the potential of an item: $\varphi_i(x) := (\text{Color}_i(x))_i(x)$ (e.g. if Color$_i(x)$ = Red then $\varphi_i(x)$ = Red$_i(x)$). Colors are assigned to elements as follows:

**Internal.** If $x$ is $i$-internal: Color$_i(x)$ = Red.
**Glancing.** If $x$ is $i$-glancing: if Red$_i(x)$ < Purple$_i(x)$ then Color$_i(x)$ = Red otherwise Color$_i(x)$ = Purple.
**External.** If $x$ is $i$-external: if Mra$_i(x).time \neq i$ then if Green$_i(x) \leq$ Blue$_i(x)$ then Color$_i(x)$ = Green, otherwise Color$_i(x)$ = Blue. If Mra$_i(x).time = i$ then Color$_i(x)$ = Blue; at any time there are one or two nodes for which this case applies, the extreme leftmost and rightmost accesses at time $i$, which are referred to as *top corners*.

Let Row$_i^{\text{Red}}$, Row$_i^{\text{Green}}$, Row$_i^{\text{Blue}}$, Row$_i^{\text{Purple}}$ be the items accessed at time $i$ with the given color, i.e. Row$_i^\gamma := \{x \in \text{Row}_i | \text{Color}_i(x) = \gamma\}$. Observe that the actual cost of time $i$ is the total number of accesses at time $i$, which is $|\text{Row}_i| = |\text{Row}_i^{\text{Red}}| + |\text{Row}_i^{\text{Green}}| + |\text{Row}_i^{\text{Blue}}| + |\text{Row}_i^{\text{Purple}}|$.

**Virgin potential.** Suppose $r$ is a root of $F_i$, and for some $k \leq i$, $r$ was not a root of $F_{k-1}$ and was a root of all $F_j$, $k \leq j \leq i$ (i.e., $r$ has continuously been a root from time $k$ to $i$). Root $r$ is deemed to be an *i-virgin* if there have been no accesses to any items in its interval $I_r$ in times $j \in [k, i]$, that is, since it has been a root. An $i$-virgin root has a potential of 12, while a non-$i$-virgin root has a virgin potential of zero. We denote the virgin potential of root $r$ at time $i$ as $v_i(r)$. Additionally, it will be useful to define $t_i$ to be the number of distinct roots in $F_i$ that have at least one item in their range accessed at time $i$.

**Potential.** The potential $\Phi_i$ is simply the sum of $\varphi_i(x)$ for all $i \in [1..n]$ and $v_i(r)$ for all roots in $F_i$.

For the analysis we find it useful to define an *intermediate potential* where we have updated the reference tree but not the point set. We denote the intermediate potential by $\Phi'_i$. To compute $\Phi'_i$ we set $F'_i = F_i$ but have $P'_i = P_{i-1}$. As all other notation derives from $P$ and $F$ we define primed variants $\varphi'_i$, Mra$'_i$, L-Int$'_i$, R-Int$'_i$, L-Ext$'_i$, R-Ext$'_i$, $v'_i$, $i$-external$'$, $i$-internal$'$, $i$-glancing$'$ and Color$'_i$ based on $F'_i$ and $P'_i$. As several of these that depend solely on $P'_i = P_{i-1}$ and not $F'_i = F_i$, we have: Mra$'_i$ = Mra$_{i-1}$, L-Int$'_i$ = L-Int$_{i-1}$, R-Int$'_i$ = R-Int$_{i-1}$, L-Ext$'_i$ = L-Ext$_{i-1}$, R-Ext$'_i$ = R-Ext$_{i-1}$, $i$-external$'$ iff $i-1$-external, $i$-internal$'$ iff $i-1$-internal, $i$-glancing$'$ iff $i-1$-glancing.

The colored sets Row$_i^{\text{Red}'}$, Row$_i^{\text{Green}'}$, Row$_i^{\text{Blue}'}$, and Row$_i^{\text{Purple}'}$ take on a special meaning. We define Row$_i^{\gamma'} := \{x \in \text{Row}_i | \text{Color}'_i(x) = \gamma\}$, that is, the set of elements accessed in Row$_i$, that were of color $\gamma$ after the tree was updated (but using the old point set $P'_i = P_{i-1}$ to determine the colors).

**Overview.** The potential method is used whereby the amortized cost of operation $i$ is the actual cost of operation $i$, $|\text{Row}_i|$, plus the change in potential $\Phi_i - \Phi_{i-1}$. We bound the potential change from $\Phi_{i-1}$ to the intermediate potential $\Phi'_i$ in §3.4; we call this *Phase I*. We then bound $\Phi_i - \Phi'_i$ in §3.5, this is fairly involved and has a number of cases to consider; we call this *Phase II*. Finally in §3.6 we combine the two potential bounds with the actual cost to obtain the main result.

### 3.3 Facts

This section contains a number of observations which are used in the main proofs but are generally intuitive and are moved here to improve the flow of the main proofs. All proofs that use left and right can be reversed by symmetry.

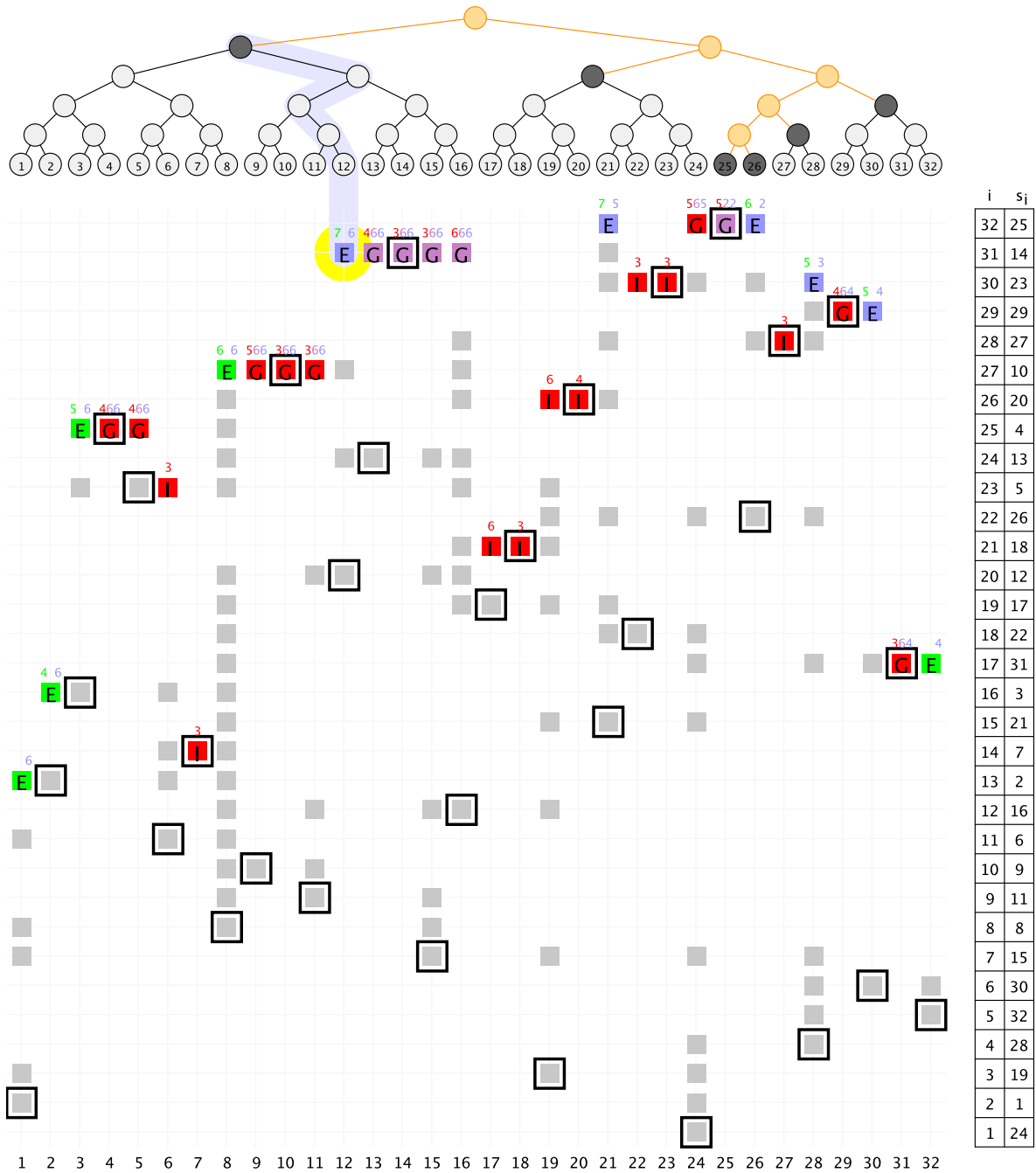

Figure 4: The computation of the blue potential of item 12 is illustrated. This is done by simply following the leaf labeled 12 up to the root of its tree in the forest and counting that five nodes are encountered and adding one to get six. Blue numbers above external and glancing items indicate their blue potential. The purple numbers above glancing items indicate the highest the blue potential has been since that item become glancing.

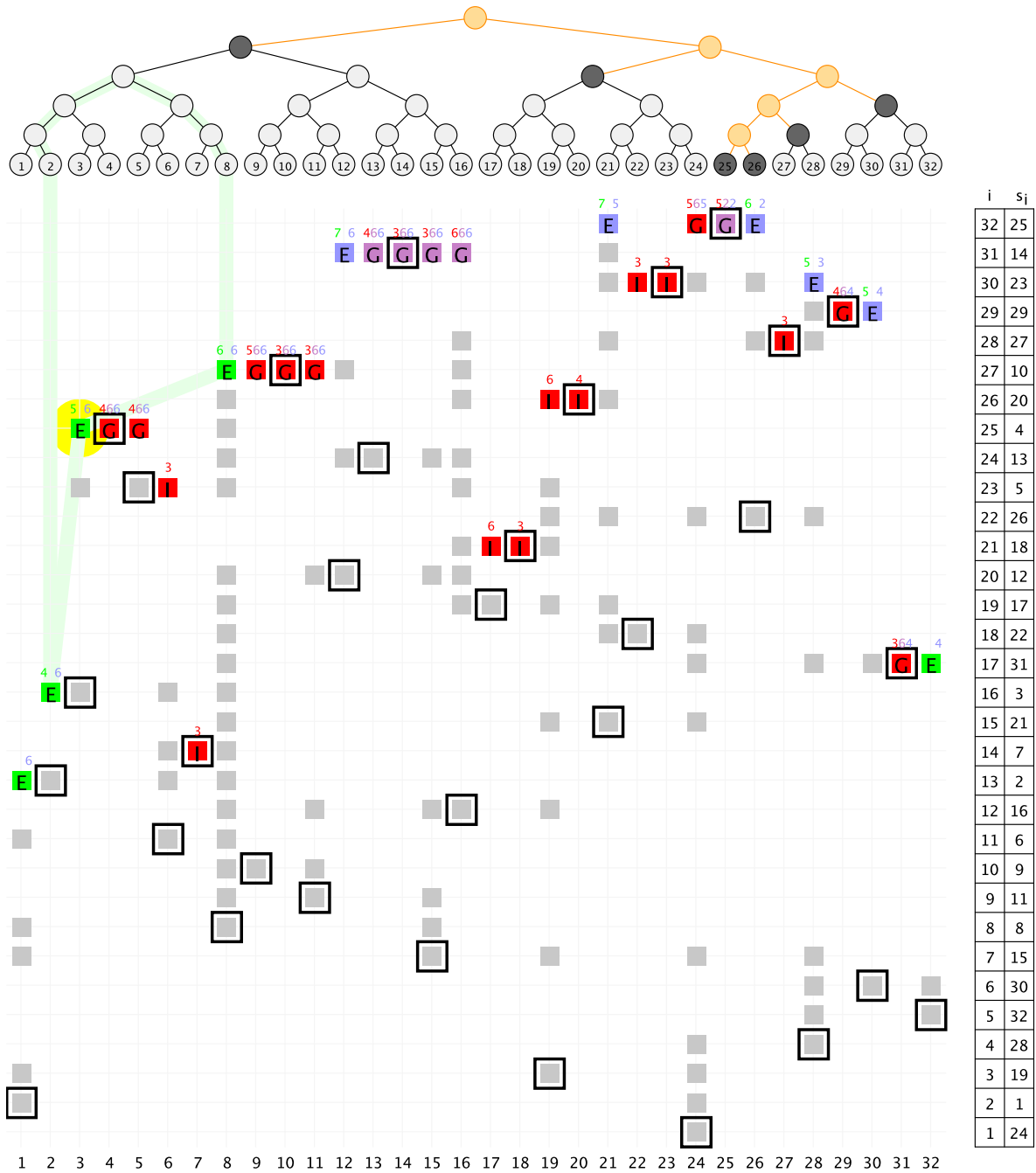

Figure 5: The computation of the green potential of item 3 is illustrated. The external items to the left and right of 3 are identified, in this case they are 2 and 8. Then the LCA of 2 and 8 is computed and is drawn in green on the figure. Finally, the green potential is one plus the number of nodes from the leaf containing item 3 to the LCA, in this case $1 + 4 = 5$.

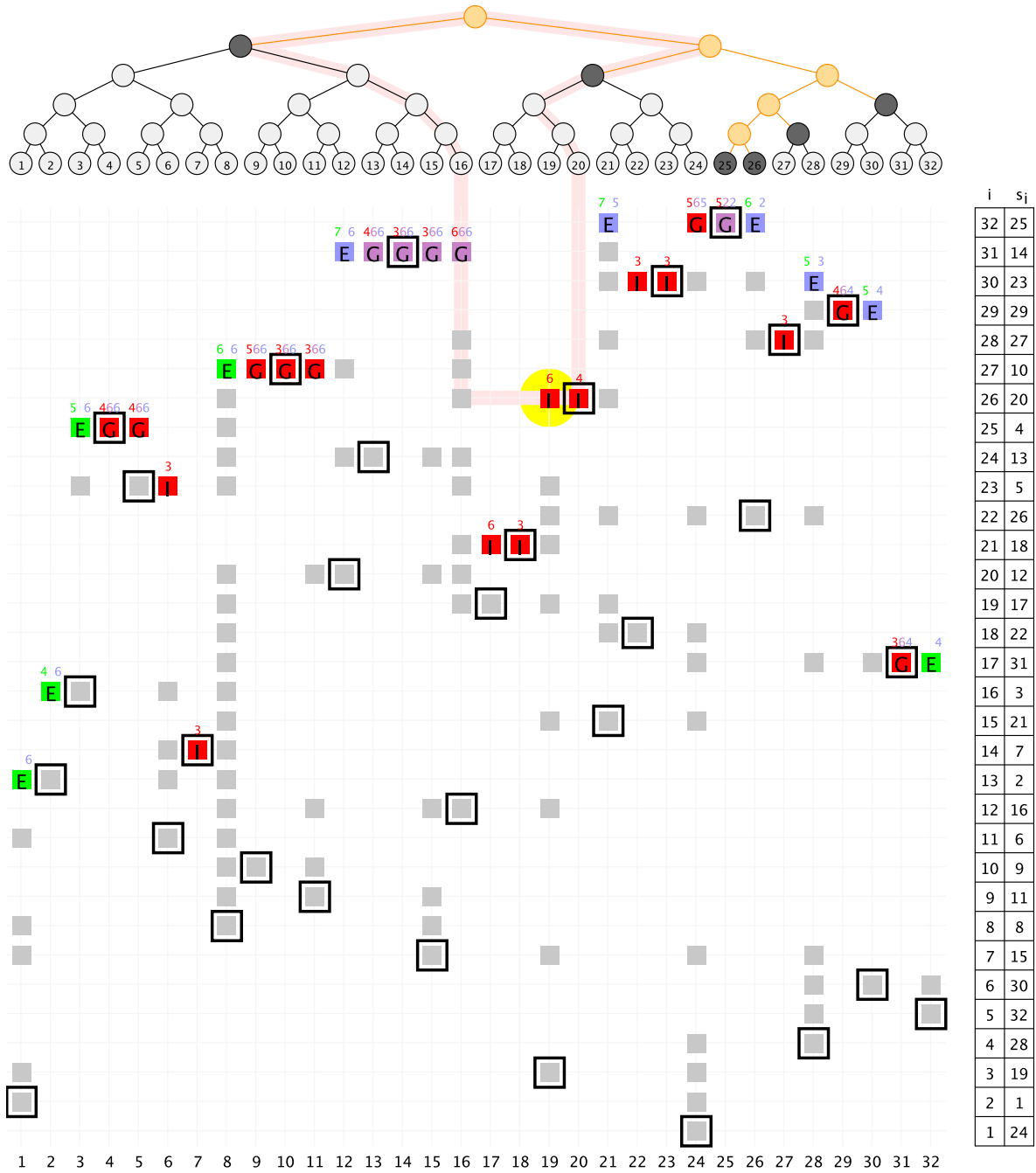

Figure 6: The computation of the red potential of 19 is illustrated. The items L-Int(19) and R-Int(19 are identified, in this case the are 16 and 20. Then the LCA of 16 and 20 is computed and is drawn in red on the figure. Finally, the red potential is the number of nodes from the leaf containing 19 to the LCA, in this case 6.

FACT 3.1. *Given leaves $w$, $x$, $y$, $z$ appearing in that order in a tree $T$, $d_T(x, \text{LCA}_T(w, y)) \leq d_T(x, \text{LCA}_T(w, z))$ and $d_T(y, \text{LCA}_T(x, z)) \leq d_T(y, \text{LCA}_T(w, z))$.*

*Proof.* This is obvious. □

FACT 3.2. *If $x$ is $i$-external, then $\text{COLOR}_i(x) = \text{BLUE}$ iff at least one of $\text{L-EXT}_i(x)$ and $\text{R-EXT}_i(x)$ is not in $T_{r_i(x)}$. Conversely, $\text{COLOR}_i(x) = \text{GREEN}$ iff both $\text{L-EXT}_i(x)$ and $\text{R-EXT}_i(x)$ are in $T_{r_i(x)}$.*

*Proof.* To see this, observe that $\text{GREEN}_i(x) \leq \text{BLUE}_i(x)$ exactly when $\text{LCA}(\text{L-EXT}_i(x), \text{R-EXT}_i(x))$ is no higher in $T$ than $r_i(x)$. □

FACT 3.3. (INVARIANCE OF LEFT INTERNAL IF NOT ACCESSED.) *If $x$ has $\text{L-INT}_{i-1}(x)$ defined, and $x \notin \text{ROW}_i$ then $\text{L-INT}_{i-1}(x) = \text{L-INT}_i(x)$.*

*Proof.* Item $\text{L-INT}_{i-1}(x)$ is defined to be the access to the left of $\text{MRA}_{i-1}(x)$. Since $x \notin \text{ROW}_i$ ($x$ is not accessed at time $i$) then $\text{MRA}_i(x) = \text{MRA}_{i-1}(x)$ which immediately gives the fact. □

### 3.4 Phase I: From the previous potential to the intermediate potential

In this step we consider the potential changes caused by updating the reference forest. To obtain a bound on this in Lemma 3.3, we first analyze the potential changes caused by small changes to the reference forest. In turn to bound this we need a lemma bounding the number of items with blue and purple potential; this Lemma 3.1 will find use as well in proof of Lemma 3.19

LEMMA 3.1. *At any time $i$, there are at most 4 blue and 2 purple items in the range of any root of $F_i$.*

*Proof.* Suppose there are more than two purple glancing items in the range of some root $r$ of $F_i$. Denote three of these items as $x, y, z$ such that $x < y < z$. Observe that:

$$\text{RED}_i(y) = d(y, \text{LCA}(\text{L-INT}_i(y), \text{R-INT}_i(y)))$$

Since $x \leq \text{L-INT}_i(y)$ and $y \geq \text{R-INT}_i(y)$, by Fact 3.1:

$$\leq d(y, \text{LCA}(x, z))$$

Since $r_i(x) = r_i(y) = r_i(z)$:

$$\leq d(y, r_i(y))$$
$$= \text{BLUE}_i(y) - 1$$
$$< \text{BLUE}_i(y)$$
$$\leq \max_{j=\text{MRA}_i(y).time}^{i} \text{BLUE}_j(y)$$
$$= \text{PURPLE}_i(y)$$

Thus since $\text{RED}_i(y) < \text{PURPLE}_i(y)$, by the definition of the color of a $i$-glancing item, $y$'s color will be RED, a contradiction.

By an similar argument, we now show there are no more than two $i$-left external items and no more than two $i$-right external items with blue potential.

Suppose there are more than two $i$-left-external blue items in the range of some root $r$ of $F_i$. Denote three of these items as $x, y, z$ such that $x < y < z$. Observe that $y$ cannot be a top corner, and:

$$\text{GREEN}_i(y) = d(y, \text{LCA}(\text{L-EXT}_i(y), \text{R-EXT}_i(y))) + 1$$

Since $x \leq \text{L-EXT}_i(y)$ and $y \geq \text{R-EXT}_i(y)$, by Fact 3.1:

$$\leq d(y, \text{LCA}(x, z)) + 1$$

Since $r_i(x) = r_i(y) = r_i(z)$:

$$\leq d(y, r_i(y)) + 1$$
$$= \text{BLUE}_i(y)$$

Thus since $\text{GREEN}_i(y) \leq \text{BLUE}_i(y)$ by the definition of the color of an $i$-external item, and because $y$ is not a top corner, $y$'s color will be GREEN, a contradiction. □

We will visualize the process of transforming the forest $F_{i-1}$ to $F_i$ as a sequence of splits and merges; see Figure 7. A *split* removes a root from the reference forest and replaces it by the subtrees of its two children (thus adding one to the number of trees in the forest), while a *merge* is the complementary operation.

LEMMA 3.2. *Splits and merges in the reference tree have a potential increase of at most* 24 *each.*

*Proof.* Changes are possible in both the virgin and item potentials. With regard to the virgin potentials, new roots are virgins. There is one new root in a merge and two in a split. The potential gain is at most 12 per new virgin.

Note that whether an item is $i$-internal, $i$-external, or $i$-glancing does not depend on $F_i$ and thus is invariant on splits and merges. The only colors that are affected by splits and merges are blue and purple, since their potential depends on the distance to their root, which may change and will only grow in a a merge. By lemma 3.1 there are at most 6 keys in the range of any root which are blue or purple. The increase in potential in each of these nodes is at most 1, as the distance to the root will increase by exactly one in a merge. Glancing items may change color from purple to red, but only if their potential was tied, and their color was therefore designated as purple, after purple increases the item becomes red and there is no potential change. External items can grange from green to blue for the same reasons, and also have no potential gain.

Thus in a merge, there is one new virgin and at most 6 blue/purple nodes that increase item potential by at most one, while in a split there are two new virgins and no increase in potential of blue/purple nodes. In either case, at most 24 units of potential are gained. □

LEMMA 3.3. (PHASE I POTENTIAL CHANGE) *The potential gain from the previous potential to the intermediate potential,* $\Phi'_i - \Phi_{i-1}$, *is at most* $24 d_T(s_{i-1}, s_i)$.

*Proof.* Recall that the only difference in how the intermediate potential $\Phi'_i$ is computed as compared to $\Phi_{i-1}$ is that the reference forest $F_i$ is used rather than $F_{i-1}$. Conceptually, we can convert $F_{i-1}$ to $F_i$ by performing a series of merges starting at the leaf with $s_{i-1}$ up to the LCA of $s_i$ and $s_{i-1}$, and then performing a series of splits from the LCA down to $s_i$. By Lemma 3.2, performing the splits and merges needed will cause an increase in potential of at most $24 d_T(s_{i-1}, s_i)$. □

## 3.5 Phase 2: From the intermediate potential to the new potential

In this phase of the analysis, we bound the changes by adding the points in the geometric view representing the items accessed at time $i$. These are geometrically added to the $i$th row in the geometric view and are located in the columns of $\text{ROW}_i$. However, due to the fact that we are beginning with the intermediate potential we can assume that the reference forest is $F_i$. Adding the accesses in the $i$th row can change the potential in many ways. Roots that were virgins may have accesses in their range and thus be virgins no more; we bound changes in virgin potential in §3.5.1. Items that are accessed at time $i$ will become glancing or top corner external, and generally incur a potential change; these changes are bounded in §3.5.2. Finally, in §3.5.3 we examine changes in potential might that be possible even for items that are not accessed.

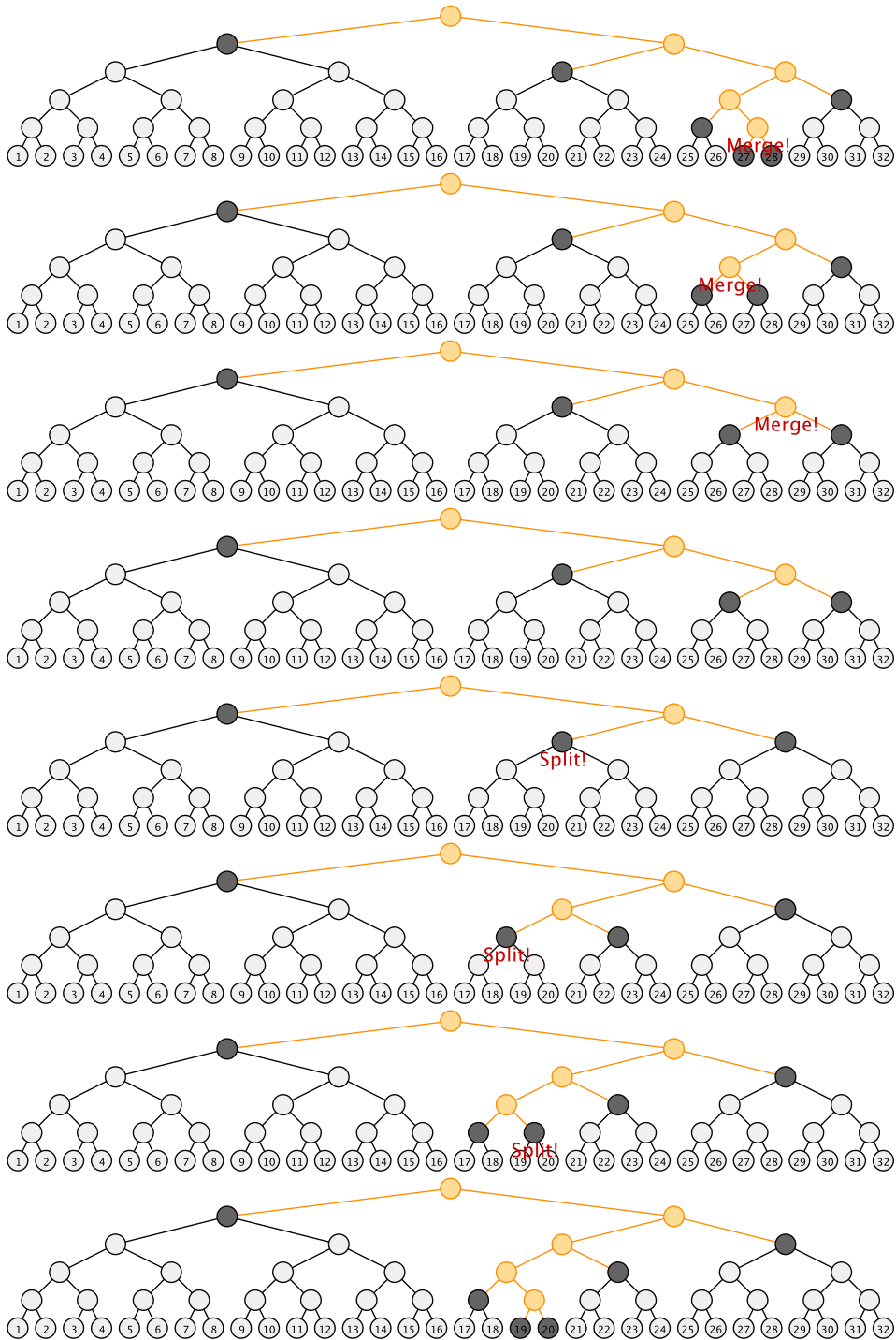

Figure 7: Illustration of the splitting and merging steps needed to turn $F_{i-1}$ into $F_i$. Here $s_{i-1}$ is 27 and $F_i$ is 19.

### 3.5.1 Virgin changes

LEMMA 3.4. *If key $x \in \text{Row}_i$, and $s_{i+1}, s_{i+2}, \ldots s_k$ all are greater than $x$, no item smaller than $x$ will be accessed at times $i+1..k$. (and symmetrically). Equivalently, if point $(x,i) \in P$, then if the rectangle $[1, x-1] \times [i+1, k]$ is free of searches, it is free of accesses.*

*Proof.* This is due to the way GreedyASS runs. Item $x$ is accessed at time $i$ iff there is a rectangle from $\text{MRA}_{i-1}(x)$ to $(s_i, i)$ that contains only $\text{MRA}_{i-1}(x)$ in $P_{i-1}$. Suppose the lemma was false, and some item $y < x$ was accessed as the first violation of the lemma, and suppose this violation happens at time $k \geq j > i$. Then consider the rectangle defined by $(s_j, j)$ and $\text{MRA}_{j-1}(y) = \text{MRA}_i(y)$. This is not empty as it contains the point $(x, i)$ and thus $y$ is not accessed, a contradiction. □

LEMMA 3.5. *If $r$ is a root of all the reference forests from time $i$ to $j$, $F_i \ldots F_j$, then all searches $s_i \ldots s_j$ are to elements not in the interval of $r$ and on the same side of the interval of $r$.*

*Proof.* Consider the following stronger statement: If $r$ is a root of the reference forest $F_i$, all searches are in the interval of its sibling (which is in $T$ but not $F_i$). This is because if $r$ is a root in $F_i$, then the parent of $r$ must be an ancestor of the current search $s_i$, but $r$ itself is not. This can only occur if the search $s_i$ is in the subtree of $r$'s sibling. □

LEMMA 3.6. *If $r$ is a non-virgin root, then there are no accesses in any tree in $F_i$ to the side of $r$ opposite the current search $s_i$.*

*Proof.* A non-virgin root has been accessed, and by Lemma 3.5 all subsequent searches are to the same side of the root as the initial search. Thus by Lemma 3.4 there can be no accesses to the opposite side. □

LEMMA 3.7. *At most two non-virgin roots have accesses in their range at any time.*

*Proof.* If more than two non-virgin roots have accesses in their range, there must be two non-virgin roots with accesses on the same side of the search. This would violate Lemma 3.6. □

LEMMA 3.8. *At least $12(t_i - 2)$ units of virgin potential are lost in Phase II.*

*Proof.* Out of the $t_i$ roots in $F_i$ that have accesses in their intervals at time $i$, at most two are non-virgins by Lemma 3.7. Those have no virgin potential to lose. The other $\geq t_i - 2$ of them were virgins and lose 12 units of potential each. The roots that do not have accesses in their intervals do not have any change of virginity. □

**3.5.2 Potential changes of accessed items** We break the analysis of the potential changes of items that are accessed into cases depending on their color at the end of Phase I. Note that, by the definition of the coloring, all accessed elements are colored RED, PURPLE, or BLUE. They cannot be GREEN since the only $i$-external elements in $\text{Row}_i$ are the two top corners.

LEMMA 3.9. (ACCESSED ITEMS THAT WERE GREEN AFTER PHASE I) *For all $x \in \text{Row}_i^{\text{GREEN}'}$ then $\varphi_i(x) - \varphi_i'(x) = -1$*

*Proof.* Suppose w.l.o.g. that $x$ is $i$-left external′. Since $\text{COLOR}_i'(x) = \text{GREEN}$, $x$ is not a top corner in $P_i'$ as those are BLUE by definition. Also, L-$\text{EXT}_i'(x)$, $x$, and R-$\text{EXT}_i'(x)$ are in the same tree in $F_i' = F_i$; otherwise $x$ would have $\text{COLOR}_i'(x) = \text{BLUE}$ by Fact 3.2. The search $s_i$ must be to the left of L-$\text{EXT}_i'(x)$, $x$, and R-$\text{EXT}_i'(x)$; $s_i$ is not in the same tree as these three and must be entirely to the left or to the right, but, if it were to the right, R-$\text{EXT}_i'(x)$ would block $x$ from being accessed. Since $s_i$ is to the left of the three external items L-$\text{EXT}_i'(x)$, $x$, and R-$\text{EXT}_i'(x)$, these three will all be accessed and no items between them will be accessed. Thus L-$\text{INT}_i(x) = $ L-$\text{EXT}_i'(x)$ and R-$\text{INT}_i(x) = $ R-$\text{EXT}_i'(x)$. So:

$$\begin{aligned}\varphi_i'(x) &= \text{GREEN}_i'(x) \\ &= d(x, \text{LCA}(\text{L-EXT}_i'(x), \text{R-EXT}_i'(x))) + 1 \\ &= d(x, \text{LCA}(\text{L-INT}_i(x), \text{R-INT}_i(x))) + 1 \\ &= \text{RED}_i(x) + 1\end{aligned}$$

If $\text{Color}_i(x) = \text{Red}$, then $\varphi_i(x) = \text{Red}_i(x) = \varphi'_i(x) - 1$ and 1 potential is lost. If $\text{Color}_i(x) = \text{Purple}$, then by the color assignment rules, $\text{Red}_i(x) \geq \text{Purple}_i(x)$, and $\varphi_i(x) = \text{Purple}_i(x) \leq \text{Red}_i(x) = \varphi'_i(x) - 1$ and one potential is lost. Finally, we argue that $\text{Color}_i(x) \neq \text{Blue}$, otherwise $x$ would be a top corner in $P_i$ and $i$-external, but this cannot happen since $s_i$ is to the left of $x$, and R-$\text{Ext}'_i(x)$ is to its right. □

LEMMA 3.10. (ACCESSED ITEMS THAT WERE PURPLE AFTER PHASE I) *For all $x \in \text{Row}_i^{\text{Purple}'}$, $\varphi_i(x) - \varphi'_i(x) \leq 0$.*

*Proof.*

$$\varphi'_i(x) = \text{Purple}'_i(x)$$

Using the definition of Purple:

$$= \max_{j=\text{Mra}'_i(x).time}^{i} \text{Blue}'_j(x)$$
$$\geq \text{Blue}'_i(x)$$

Since $\text{Blue}'_i(x) = \text{Blue}_i(x)$:

$$= \text{Blue}_i(x)$$

Since $x$ is accessed; $\text{Mra}_i(x).time = i$:

$$= \max_{j=\text{Mra}_i(x).time}^{i} \text{Blue}_j(x)$$

By the definition of Purple:

$$= \text{Purple}_i(x)$$

If $\text{Color}_i(x) = \text{Purple}$, then $\varphi_i(x) = \text{Purple}_i(x) \leq \varphi'_i(x)$. If $\text{Color}_i(x) = \text{Red}$, then by definition of the coloring, $\varphi_i(x) = \text{Red}_i(x) < \text{Purple}_i(x) \leq \varphi'_i(x)$. Finally, if $\text{Color}_i(x) = \text{Blue}$, then $\varphi_i(x) = \text{Blue}_i(x) \leq \varphi'_i(x)$ by the fourth line of this proof. □

LEMMA 3.11. (ACCESSED ITEMS THAT WERE BLUE AFTER PHASE I) *For all $x \in \text{Row}_i^{\text{Blue}'}$, $\varphi_i(x) - \varphi'_i(x) \leq 0$.*

*Proof.*

$$\varphi'_i(x) = \text{Blue}'_i(x)$$

Since $\text{Blue}'_i(x) = \text{Blue}_i(x)$:

$$= \text{Blue}_i(x)$$

Since $x$ is accessed and $\text{Mra}_i(x).time = 1$:

$$= \max_{j=\text{Mra}_i(x).time}^{i} \text{Blue}_j(x)$$

By the definition of Purple:

$$= \text{Purple}_i(x)$$

If $\text{Color}_i(x) = \text{Purple}$, then $\varphi_i(x) = \text{Purple}_i(x) = \varphi'_i(x)$. If $\text{Color}_i(x) = \text{Red}$, then by definition of the coloring, $\varphi_i(x) = \text{Red}_i(x) \leq \text{Purple}_i(x) - 1 = \varphi'_i(x) - 1$. Finally, if $\text{Color}_i(x) = \text{Blue}$, then $\varphi_i(x) = \text{Blue}_i(x) = \varphi'_i(x)$ by the second line of this proof. □

LEMMA 3.12. (ACCESSED ITEMS THAT WERE RED AFTER PHASE I) *For all* $x \in \text{Row}_i^{\text{Red}'}$, $\varphi_i(x) - \varphi'_i(x) \leq -1$ *if* $\text{Color}_i(x) = \text{Red}$ *and* $\varphi_i(x) - \varphi'_i(x) \leq 0$ *otherwise.*

*Proof.* Assume w.l.o.g. that $x < s_i$.

$$\varphi'_i(x) = \text{Red}'_i(x)$$

By the definition of $\text{Red}_i(x)$:

$$= d(x, \text{Lca}(\text{L-Int}'_i(x), \text{R-Int}'_i(x)))$$

Observe that $\text{R-Int}'_i(x)$ is to the right of $s_i$; otherwise $x$ would not have been accessed by GreedyASS as $\text{Mra}'_i(\text{R-Int}'_i(x))$ would lie in the rectangle formed by $s_i$ and $\text{Mra}'_i(x)$. Thus, by Fact 3.1 and $\text{R-Int}'_i(x) \geq s_i$:

$$\geq d(x, \text{Lca}(\text{L-Int}'_i(x), s_i))$$

Using Fact 3.1 and $\text{L-Int}'_i(x) < x$:

$$\geq d(x, \text{Lca}(x, s_i))$$

Observe that since $s_i$ is not in the same tree of $F_i$ as $x$, the node $LCA(x, s_i)$ in $T$ is above the root $r_i(x)$ in $F_i$ and thus $LCA(x, s_i) \geq d(x, r'_i(x)) + 1$.

$$\geq d(x, r'_i(x)) + 1$$

Since $F_i = F'_i$ and thus $r'_i(x) = r_i(x)$

$$\geq d(x, r_i(x)) + 1$$

By the definition of BLUE:

$$= \text{Blue}_i(x)$$

Since $\text{Mra}_i(x).time = i$:

$$= \max_{j=\text{Mra}_i(x).time}^{i} \text{Blue}_j(x)$$

$$= \text{Purple}_i(x)$$

Thus, if $\varphi_i(x) = \text{Blue}_i(x)$, $\varphi_i(x) - \varphi'_i(x) \leq 0$ and if $\varphi_i(x) = \text{Purple}_i(x)$, $\varphi_i(x) - \varphi'_i(x) \leq 0$. This leaves the case $\text{Color}_i(x) = \text{Red}$. But then $\text{Red}_i(x) < \text{Purple}_i(x)$ so $\varphi_i(x) = \text{Red}_i(x) \leq \text{Purple}_i(x) - 1 \leq \varphi'_i(x) - 1$, so $\varphi_i(x) - \varphi'_i(x) \leq -1$ holds in this case as well. □

### 3.5.3 Potential changes of unaccessed items
Here we consider the possible potential changes of those items that are not accessed.

LEMMA 3.13. (UNACCESSED ITEMS THAT WERE EXTERNAL (GREEN OR BLUE) AFTER PHASE I) *For all* $x \notin \text{Row}_i$, *if* $x$ *is* $i$*-external'*, *then* $\varphi_i(x) - \varphi'_i(x) \leq 0$.

*Proof.* First consider the case when $x$ is not a top corner. Assume w.l.o.g that $x$ is $i$-left external'. Item $x$ is not in $\text{Row}_i$ and thus does not become a top corner. If $s_i < x$ then $x$ would have been accessed since being external means that the rectangle from $s_i$ to $\text{Mra}_i(x)$ is empty, and $s_i \neq x$, thus $s_i > x$. No item to the left of $x$ is accessed; such an access would cause an ASS violation with $\text{Mra}_i(x)$; thus $x$ remains external and with the same external neighbor to the left $\text{L-Ext}_i(x) = \text{L-Ext}'_i(x)$. If $s_i > \text{R-Ext}'_i(x)$ then no items to the left of $\text{R-Ext}'_i(x)$ are accessed, R-EXT$_i(x)$ remains external and R-EXT$_i(x)$ = R-EXT$'_i(x)$. Else, R-EXT$_i(x)$ < R-EXT$'_i(x)$; in either case R-EXT$_i(x) \leq$ R-EXT$'_i(x)$. Using these observations gives the claim after unrolling and re-rolling the definitions:

By the definition of the potential of a non-corner external:

$$\varphi'_i(x) = \min(\text{GREEN}'_i(x), \text{BLUE}'_i(x))$$

Since BLUE$_i(x)$ = BLUE$'_i(x)$:

$$= \min(\text{GREEN}'_i(x), \text{BLUE}_i(x))$$

By the definition of green potential

$$= \min(d(x, \text{LCA}(\text{L-EXT}'_i(x), \text{R-EXT}'_i(x))) + 1,$$
$$\text{BLUE}_i(x))$$

Since L-EXT$_i(x)$ = L-EXT$'_i(x)$:

$$= \min(d(x, \text{LCA}(\text{L-EXT}_i(x), \text{R-EXT}'_i(x))) + 1,$$
$$\text{BLUE}_i(x))$$

By Fact 3.1 and R-EXT$_i(x) \leq$ R-EXT$'_i(x)$

$$\geq \min(d(x, \text{LCA}(\text{L-EXT}_i(x), \text{R-EXT}_i(x))) + 1,$$
$$\text{BLUE}_i(x))$$

Using the definition of green potential:

$$= \min(\text{GREEN}_i(x), \text{BLUE}_i(x))$$

By the definition of the potential of a non-corner external

$$= \varphi_i(x)$$

When $x$ is a top corner, then

By the definition of the potential of a corner external:

$$\varphi'_i(x) = \text{BLUE}'_i(x)$$

Since BLUE$_i(x)$ = BLUE$'_i(x)$

$$= \text{BLUE}_i(x)$$

Since min's don't make things larger:

$$\geq \min(\text{GREEN}_i(x), \text{BLUE}_i(x))$$

By the definition of the potential of a non-corner external:

$$= \varphi_i(x).$$

□

LEMMA 3.14. (RED POTENTIAL OF UNACCESSED ITEMS) *For all $x \notin \text{Row}_i$ that are not i-external', $\text{Red}_i(x) = \text{Red}'_i(x)$.*

*Proof.* By the definition of red potential:

$$\text{Red}'_i(x) = d(x, \text{Lca}(\text{L-Int}'_{i-1}(x), \text{R-Int}'_{i-1}(x)))$$

Since L-INT and R-INT don't change in Phase I:

$$= d(x, \text{Lca}(\text{L-Int}_{i-1}(x), \text{R-Int}_{i-1}(x)))$$

By Fact 3.3

$$= d(x, \text{Lca}(\text{L-Int}_i(x), \text{R-Int}_i(x)))$$

By the definition of red potential:

$$= \text{Red}_i(x)$$

$\square$

LEMMA 3.15. (PURPLE POTENTIAL OF UNACCESSED ITEMS) *For all $x \notin \text{Row}_i$ that are i-glancing', $\text{Purple}_i(x) = \text{Purple}'_i(x)$.*

*Proof.* By the definition of purple:

$$\text{Purple}'_i(x) = \max_{j=\text{Mra}'_i(x).time}^{i} \text{Blue}'_j(x)$$

Since $\text{Blue}'_i = \text{Blue}_i$:

$$= \max_{j=\text{Mra}'_i(x).time}^{i} \text{Blue}_j(x)$$
$$= \text{Purple}_i(x)$$

$\square$

LEMMA 3.16. (UNACCESSED i-GLANCING' BECOMES i-INTERNAL) *If $x \notin \text{Row}_i$ is i-glancing' and i-internal, then $\text{Purple}'_i(x) \geq \text{Red}_i(x)$.*

*Proof.* The key observation is that, in order for $x$ to become glancing, there must have been searches to the right and left of $x$, by Lemma 3.4 and the definition of glancing. Let $s_l = \max\{s_j \in S | s_j < x, j \in [\text{Mra}_i(x) + 1, i]\}$ be the closest search left of $x$, and let $s_k = \min\{s_j \in S | s_j > x, j \in [\text{Mra}_i(x) + 1, i]\}$ the closest search right of $x$. Let $r$ be $\text{Lca}(s_l, s_k)$, let $r'$ be the left child of $r$ and $r''$ its right child. Item $x$ is in $I_r$ and either in $I_{r'}$ or in $I_{r''}$. Assume w.l.o.g. that it is in $I_{r'}$. Observe that $s_k$ is in $I_r$ but not in $I_{r'}$. Thus $r'$ was a root in $F_k$, and $r_k(x) = r'$. With these observations the potential can be bounded as follows:

By Lemma 3.15:

$$\text{Purple}'_i(x) = \text{Purple}_i(x)$$
$$= \max_{j=\text{Mra}_i(x).time}^{i} \text{Blue}_j(x)$$

Since $\text{Mra}_i(x).time \leq k \leq i$:

$$\geq \text{Blue}_k(x)$$

By the definition of BLUE:

$$= 1 + d(x, r_k(x))$$

Since $r_k(x) = r'$:

$$= 1 + d(x, r')$$

Since $r = \text{LCA}(s_l, s_k)$ is the parent of $r'$:

$$= d(x, \text{LCA}(s_l, s_k))$$

We know that $s_l \leq \text{L-INT}_i(x)$ and symmetrically $s_r \geq \text{R-INT}_i(x)$ as any accesses (including a search) between $x$ and $\text{L-INT}_i(x)$ at times after $\text{MRA}_i(x)$ will cause an ASS violation. Thus, using Fact 3.1:

$$\geq d(x, \text{LCA}(\text{L-INT}_i(x), \text{R-INT}_i(x)))$$

By the definition of RED:

$$= \text{RED}_i(x)$$

$\square$

LEMMA 3.17. (UNACCESSED ITEMS THAT WERE NOT EXTERNAL (PURPLE OR RED) AFTER PHASE I) *For all $x \notin \text{ROW}_i$, if $x$ is not $i$-external', then $\varphi_i(x) - \varphi'_i(x) \leq 0$.*

*Proof.* An unaccessed element $x$ that is $i$-internal' stays $i$-internal, its color is always RED, and the red potential doesn't change by Lemma 3.14, therefore its potential stays the same. For an unaccessed element $x$ that is $i$-glancing' and stays $i$-glancing, the red and purple potentials do not change by Lemmas 3.14 and 3.15, therefore the order between the two potentials, and therefore the color of $x$ and its potential stay identical. Finally, for an unaccessed element $x$ that is $i$-glancing' and becomes $i$-internal, if $\text{COLOR}'_i(x) = \text{RED}$, then it stays RED and its potential doesn't change, and if $\text{COLOR}'_i(x) = \text{PURPLE}$ (and $\text{COLOR}_i(x) = \text{RED}$ since it becomes internal), then $\varphi_i(x) - \varphi'_i(x) = \text{RED}_i(x) - \text{PURPLE}'(x) \leq 0$ by Lemma 3.16. $\square$

### 3.5.4 Bringing it all together

LEMMA 3.18. *The potential change $\Phi_i - \Phi'_i$ is at most $-12(t_i - 2) - |\text{ROW}_i^{\text{GREEN}'}| - |\text{ROW}_i^{\text{RED}'}| + |\text{ROW}_i^{\text{BLUE}}| + |\text{ROW}_i^{\text{PURPLE}}|$.*

*Proof.* We now combine the previous lemmata which together consider all possible potential changes. By Lemma 3.8, the virgin potential has a gain of at most $-12(t_i - 2)$. Lemmata 3.9-3.12 bound the potential changes of the accessed nodes depending on their color. By Lemma 3.9, any item in $\text{ROW}_i^{\text{GREEN}'}$ will drop 1 potential, and by Lemma 3.12, so will items in $\text{ROW}_i^{\text{RED}'} \cap \text{ROW}_i^{\text{RED}}$. All other accessed or non accessed nodes do not increase potential by Lemmata 3.10, 3.11, 3.13 and 3.17. Since $|\text{ROW}_i^{\text{RED}'} \cap \text{ROW}_i^{\text{RED}}| \geq |\text{ROW}_i^{\text{RED}'}| - |\text{ROW}_i^{\text{BLUE}}| - |\text{ROW}_i^{\text{PURPLE}}|$, we obtain a potential increase of at most $-|\text{ROW}_i^{\text{GREEN}'}| - |\text{ROW}_i^{\text{RED}'}| + |\text{ROW}_i^{\text{BLUE}}| + |\text{ROW}_i^{\text{PURPLE}}|$. $\square$

## 3.6 Main result

LEMMA 3.19. *The amortized cost of search $s_i$ is $24 d_T(s_{i-1}, s_i) + 24$.*

*Proof.* Recall that, by Lemma 3.1, $|\text{ROW}_i^{\text{BLUE}'}| \leq 4t_i$, $|\text{ROW}_i^{\text{BLUE}}| \leq 4t_i$, $|\text{ROW}_i^{\text{PURPLE}'}| \leq 2t_i$, and $|\text{ROW}_i^{\text{PURPLE}}| \leq 2t_i$. Using this, we can bound the amortized cost of a search.

By the definition of amortized cost

$$\hat{a}_i = |\text{Row}_i| + \Phi_i - \Phi_{i-1}$$
$$= |\text{Row}_i^{\text{RED}'}| + |\text{Row}_i^{\text{GREEN}'}| + |\text{Row}_i^{\text{BLUE}'}|$$
$$+ |\text{Row}_i^{\text{PURPLE}'}| + \Phi_i - \Phi_{i-1}$$

Adding zero:

$$= |\text{Row}_i^{\text{RED}'}| + |\text{Row}_i^{\text{GREEN}'}| + |\text{Row}_i^{\text{BLUE}'}|$$
$$+ |\text{Row}_i^{\text{PURPLE}'}| + (\Phi_i - \Phi_i') + (\Phi_i' - \Phi_{i-1})$$

By Lemmata 3.3 and 3.18:

$$= |\text{Row}_i^{\text{RED}'}| + |\text{Row}_i^{\text{GREEN}'}| + |\text{Row}_i^{\text{BLUE}'}|$$
$$+ |\text{Row}_i^{\text{PURPLE}'}| + 24 d_T(s_{i-1}, s_i)$$
$$+ (-12(t_i - 2) - |\text{Row}_i^{\text{RED}'}|$$
$$- |\text{Row}_i^{\text{GREEN}'}| + |\text{Row}_i^{\text{BLUE}}| + |\text{Row}_i^{\text{PURPLE}}|)$$
$$= |\text{Row}_i^{\text{BLUE}'}| + |\text{Row}_i^{\text{BLUE}}|$$
$$+ |\text{Row}_i^{\text{PURPLE}'}| + |\text{Row}_i^{\text{PURPLE}}|$$
$$- 12 t_i + 24 d_T(s_{i-1}, s_i) + 24$$

By Lemma 3.1:

$$= 24 d_T(s_{i-1}, s_i) + 24$$

□

THEOREM 3.1. *Let $\mathcal{T}$ be the set of all leaf-oriented BSTs containing the integers $[1..n]$ as leaves. The cost to execute search sequence $s_1, s_2, \ldots s_m$ using the BST algorithm GreedyASS is*

$$O\left(\min_{T \in \mathcal{T}} \sum_{i=1}^m d_T(s_{i-1}, s_i) + n\right).$$

*Proof.* By the potential method, we can bound the runtime of the sequence of searches by summing up the amortized cost of Lemma 3.19 along with the potential gain. As the initial potential is zero, and is always positive, there is no gain. Finally the use of the conversion in [6] to convert the geometric view of GreedyASS into a BST introduces an additional $O(n)$ cost, which is asymptotically insignificant for $m = \Omega(n)$. □

## 4 Implications

Theorem 3.1, as stated gives the lazy-finger bound with respect to a leaf-oriented reference tree. This simplified the proof. However, the lazy finger was originally presented in terms of a reference tree which was a standard static binary search tree. We now show how our result can easily be made to hold with a normal binary search tree as the reference tree, and then state the immediate consequences of this from [2].

LEMMA 4.1. *For any n-node BST $T$ containing the integers $[1..n]$, there is an $2n - 1$ node leaf-oriented BST $T'$ with leaves labeled $[1..n]$ such that $d_T(x, y) \leq 2 d_{T'}(x, y)$ for any $x, y \in [1..n]$.*

*Proof.* Replace each node $x$ of $T$ with three nodes as in figure 8 to obtain $T''$. By contracting all single-child nodes (which contain no data), the tree $T'$ with $2n - 1$ nodes is obtained. □

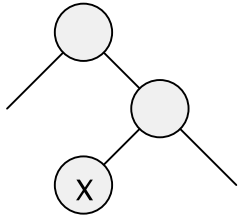

Figure 8: To convert a non-leaf oriented BST tree into a tree where all the items are in the leaves, replace every node containing item $x$ with the three nodes illustrated.

Thus combining Lemma 4.1 and Theorem 3.1 gives:

THEOREM 4.1. (LAZY FINGER THEOREM) *Let $\mathcal{T}$ be the set of all BSTs containing the integers $[1..n]$. The cost to execute a sequences of searches $s_1, s_2, \ldots s_m$ using the BST algorithm GreedyASS is*

$$O\left(n + \min_{T \in \mathcal{T}} \sum_{i=1}^{m} d_T(s_{i-1}, s_i)\right)$$

This can be equivalently stated, using the central result of [2] as:

THEOREM 4.2. (WEIGHTED DYNAMIC FINGER THEOREM) *Let $\mathcal{W}$ be the set of all $W = w_1, \ldots w_n$, with all $w_i$ positive. The cost to execute a sequences of searches $s_1, s_2, \ldots s_m$ using the BST algorithm GreedyASS is*

$$O\left(n + \min_{W \in \mathcal{W}} \left\{\sum_{i=1}^{m} \log \frac{\sum_{k=\min(s_i, s_{i-1})}^{\max(s_i, s_{i-1})} w_k}{\min(w_{s_{i-1}}, w_{s_i})}\right\}\right).$$

Since this theorem holds for the choice of weights $W$ that minimizes the runtime, it also holds for any choice of $W$. Setting all $w_i = 1$ gives:

COROLLARY 1. (DYNAMIC FINGER THEOREM) *The cost to execute a sequences of searches $s_1, s_2, \ldots s_m$ using the BST algorithm GreedyASS is $O(n + \sum_{i=2}^{m} \log |1 + s_{i-1} - s_i|)$*

Using our result on deamortizing [1] and combining [8] BST algorithms to combine GreedyASS with Tango trees, there is a BST algorithm with weighted dynamic finger, working set, is within a $O(\log \log n)$ factor of optimal, and has a $O(\log n)$ worst-case runtime:

THEOREM 4.3. (COMBINING PREVIOUS RESULTS) *Let $\mathcal{W}$ be the set of all $W = w_1, \ldots w_n$, with all $w_i$ positive. There is a BST algorithm that the cost to execute a sequences of searches $s_1, s_2, \ldots s_m$*

$$O\left(n + \min\left\{\min_{W \in \mathcal{W}} \left\{\sum_{i=1}^{m} \log \frac{\sum_{k=\min(s_i, s_{i-1})}^{\max(s_i, s_{i-1})} w_k}{\min(w_{s_{i-1}}, w_{s_i})}\right\},\right.\right.$$
$$\left.\left. OPT(S) \log \log n \right\}\right)$$

*and where each search runs in time $O(\log n)$ worst-case.*

## 5 Further work

There are a number of possible directions for further work based on the techniques presented here. First, one could attempt to extend it to where the reference tree has a constant number of pointers; this would give a stronger bound on some sequences. Secondly, one could alter the argument to make it work on Splay trees (and perhaps larger classes of BSTs such as those defined in [19, 3] ). The existing potential function does not work for Splay trees, but we believe that a modification of it will. Finally, if we were to allow arbitrary rotations in the reference tree with only constant potential change, this would prove dynamic optimality for GreedyASS. This also does not work with the current potential function but we conjecture a variant of the potential with the proof technique presented here will work.